\documentclass[a4paper,showkeys,floatfix,aps,pre,longbibliography,superscriptaddress,reprint]{revtex4-1}
\usepackage{graphics,graphicx}
\usepackage{amsmath,amssymb}

\usepackage{graphics,graphicx}
\usepackage{dcolumn,bm}
\usepackage{psfrag}
\usepackage{xstring}
\usepackage{color}
\usepackage{hyperref}
\usepackage{multirow}
\newcommand{\srm}
{\affiliation{Department of Physics, SRM University - AP, Amaravati,
 Andhra Pradesh - 522240, India}}

\newcommand{\isi}
{\affiliation{Indian Statistical Institute, Kolkata 700108, India.}}

 \newcommand{\saha}
 {\affiliation{Saha Institute of Nuclear Physics, Kolkata
700064, India}}

\begin{document}
\title[Annealing of SK and 2D-EA models]{Estimates of ground state energies for the quantum SK and 2D-EA models, using
deGennes-Suzuki-Kubo mean-field annealing dynamics}

\author{Soumyaditya Das}

\email{soumyaditya\_das@srmap.edu.in}
\srm
\author{Soumyajyoti Biswas}

\email{soumyajyoti.b@srmap.edu.in}
\srm

\author{Bikas K. Chakrabarti}
\email{bikask.chakrabarti@saha.ac.in}
\saha
\isi

\begin{abstract}
 We perform large scale quantum annealing of the Sherrington-Kirkpatrick (SK) spin glass up to a system size $N=40000$ to estimate its ground state energy using the deGennes-Suzuki-Kubo mean-field quantum Ising dynamics, extending the earlier results (reported in Eur. Phys.
J. B {\bf 98}, 226 (2025)). Here we numerically solve the deGennes-Suzuki-Kubo annealing dynamics to obtain the spin configurations and subsequently the ground state
energy for a given system size at the end of the annealing, starting
from a quantum paramagnetic state. The method shows high efficiency, with an overall algorithmic cost of $O(N^3)$ in estimating the energy of the ground state. We later extend this quantum annealing study to estimate the ground
state energies (starting again from the quantum paramagnetic phase,
annealing down to any desired low value of the transverse field) for  the
Edwards-Anderson (EA) spin glass model on a square lattice.
%We later extend this method to study the ground state energy of the Edwards-Anderson (EA) spin glass on a square lattice.   
\end{abstract}

\maketitle

\section{Introduction}
Spin glass systems, with both infinite range \cite{sher} and nearest neighbor interacting range \cite{edwards}, are paradigmatic examples of optimization problems \cite{mezard_book}, having wide range of applications from materials to computer sciences \cite{binder}. Estimating the ground state and the energy of  the ground state in many variants of the problem remain prohibitively difficult, and often recognized as NP-hard. 

The Sherrington-Kirkpatrick (SK) model and the Edwards-Anderson (EA) model are canonical versions of Ising spin glasses in the mean field and nearest neighbor interactions respectively \cite{sher,edwards}. Both have a large set of local minima in their free energy landscape, making the evolution towards the global ground state difficult. This has resulted in a wide ranging effort towards developing algorithms that can best approximate the ground state and its energy. For the SK model, such estimates can then be compared with the full Replica-Symmetry-Breaking (RSB) solution \cite{parisi1,parisi2} that yields an energy per spin value $0.7631667265 \dots$ \cite{oppermann} for the model. Such efforts to estimate the ground states and its energies include (but not limited to) simulated annealing \cite{kirk}, branch-and-bound, extremal optimization \cite{boe_epjb2005,stefan2}, continuous nonlinear optimization \cite{duxbury}, quantum annealing \cite{ray,DasChakrabarti2008} etc. 

It has been shown before that the Suzuki-Kubo-type mean field dynamical equations \cite{suzuki}, along with a Thouless-Anderson-Palmer (TAP) reaction field \cite{thouless}, when applied to the local (thermal averaged) spin variables in the SK model, can help estimate the ground state and energy in a remarkably short time ($O(N^3)$; $N$ being the system size) \cite{das1,das2}. It was argued that the local spins, when treated like a thermal averaged continuous variable can, in effect, reduce the corrugation of the free energy near the paramagnetic state. As the annealing temperature is slowly reduced, the state of the system can quickly navigate towards the global minimum, while the average spin magnitudes and thereby the barrier heights in the free energy landscape gradually increase. 

As for the quantum variant of the annealing process, where a non-commuting quantum fluctuation is usually utilized for escaping the local minima traps, the Suzuki-Kubo equations can again be utilized along with a Brout-Muller-Thomas dynamics \cite{deGennes1963,Brout1966} that results in a similar estimate of the ground state and energy, around the same time scale. The overall computational complexity, therefore, remains $O(N^3)$ \cite{das2}. The usual connection between this mean field framework and quantum spin models arises from the BCS model where Anderson \cite{Anderson1958} showed that the BCS model can be mapped onto an effective transverse-field Ising model by projecting the electronic degrees of freedom onto pair-occupied and pair-unoccupied states near the Fermi level. This mapping provides an extension of the dynamical framework developed in Refs. \cite{deGennes1963,Brout1966} beyond the BCS type model to disordered spin systems such as the SK and EA models. Although Bose-Einstein type entanglement and coherence have been studied for BCS states \cite{Peronaci}, here such quantities have not yet been systematically investigated for the quantum SK or quantum EA models.

%It was shown in ref. \cite{Anderson1958} that all the BCS model results could be obtained from an effective transverse field Ising model by projecting the electronic states to the pair occupied-unoccupied states near the Fermi-level. Here, following refs. \cite{deGennes1963,Brout1966}, we extend the concept to the dynamics of spin glasses. The studies on the Bose-Einstein type entanglement and coherence have been studied for the BCS states \cite{Peronaci}, but we have not explored them here as no study has been reported so far for the quantum SK or the quantum EA models to compare.

While quantum fluctuations do not appear to provide any additional advantage in finding the ground state of the SK Hamiltonian, it is nevertheless useful to formulate the problem within the quantum annealing framework, since the transverse field dynamics provides a unified protocol that can be applied to both classical and quantum spin glass Hamiltonians. Thus, the SK model serves as a benchmark for the performance of the deGennes-Suzuki-Kubo dynamics, while its application to the EA model allows us to examine the same framework when a finite quantum component is retained after annealing.

Here we estimate the ground state energies of  the classical SK model \cite{sher}, as well as the classical and quantum EA models \cite{edwards}, employing the zero temperature deGennes-Suzuki-Kubo mean-field quantum annealing dynamics \cite{das2,das3} for the SK model in transverse field \cite{ishii} and the EA model in transverse field \cite{bkc1981}. We show that the annealing dynamics enables us to accurately estimate the ground state energy of the SK model up to $N=40000$ (largest size studied so far, to our knowledge). We extended the same study for estimating the ground state energy of the 2D-EA model up to a system size $N=128\times128$.  Furthermore, the EA model studied here also yields ground state energy estimates for annealing down to small values of transverse fields and
the results match well with other more sophisticated estimates \cite{coffrin,pilati2024}. It is noteworthy that the Suzuki-Kubo dynamics, which is essentially for mean field interactions, works here reasonably well even for short range interactions of the EA spin glass. These results suggest that this framework may provide a useful approach for estimating ground state energy of quantum many-body systems. %This opens up the possibility of using this framework for finding the ground states of quantum many body systems. 

\section{deGennes-Suzuki-Kubo dynamics}
The Hamiltonian of the SK model 
%(for a particular realization [c] of the disorder $J_{ij}$)
in the presence of a transverse field $\Gamma$ reads
    \begin{equation}
 H^{} = - \sum_{i < j} J_{ij}\sigma_i^z\sigma_j^z
 -\Gamma \sum_i \sigma_i^x,
\label{sk_h}
\end{equation}

\noindent where $\sigma^{z,x}$ denote the Pauli
spin matrices and the coupling 
$J_{ij}$ are the quenched random variables, which specify the long-range (random ferromagnetic or antiferromagnetic) interaction between the $i$ -th and
$j$ -th spins in the SK model with Gaussian distribution centered at zero:

\begin{equation}
P (J_{ij}) = (1/J)(N/2\pi)^{1/2}
\exp[-(N/2)(J_{ij}/J)^2],       
\end{equation}
\noindent with 
\begin{equation}
[J_{ij}^2]_{av} - [J_{ij}]_{av}^2
= J^2/N = 1/N.
\end{equation}

However, one can write a simplified version of the Hamiltonian (Eq. [\ref{sk_h}]) where an effective field $\vec{h}^{\text{eff}}$ now acts on the coarse-grained spins $\vec{m}^{[c]}(=<\vec{\sigma}^{[c]}>)$ under the mean-field approximation (see for example \cite{AcharyyaPhysicaA1994,AcharyyaJPhysA1994}). Then the mean-field Hamiltonian (for a particular realization [c] of disorder $J_{ij}$) can be written as,

\begin{equation}
 H^{[c]}= - \sum_i  \vec{h}_i^{\text{eff}\,[c]}\cdot\vec{m_i}^{[c]}. 
\end{equation}

The effective field $\vec{h}^{\text{eff}}$ has two components, one is Curie-Wiess type cooperative interaction among spins of the z-component corrected by the modified Thouless-Anderson-Palmer (TAP) reaction field $(1-q)m^z$ (see refs. \cite{das2,das3} for the justification of modified TAP field)and the other is due to the external transverse field ($\Gamma$):

\begin{equation}
   {\vec{h}_i^{\text{eff}\,[c]}}= {{h}^{z\,\text{eff}\,[c]}_i}\hat{z}+{h_i}^{x\,\text{eff}\,[c]}\hat{x},
\end{equation}
\noindent where
%\begin{widetext}
\begin{equation}
\left|\vec{h}_i^{\text{eff}\,[c]}\right| = \left[ \left( \sum_j J_{ij} m_j^{z\,[c]} -\left(1 - q^{[c]}\right)\, m_i^{z\,[c]} \right)^2 + \Gamma^2 \right]^{1/2}  \tag{5a}
 %\vec{{{h'_i}^{eff}}^{[c]}} =\left[\left(\sum_j J_{ij} {{m_j}^z}^{[c]}+ (1 - {q^z}^{[c]}){{m_i}^z}^{[c]}}\right)^2 + h^2\right]^{1/2},   
\end{equation}
%\end{widetext}
\noindent with 
\begin{equation}
   h_i^{z\,\text{eff}\,[c]}= \sum_j J_{ij} m_j^{z\,[c]} - \left(1 - q^{[c]}\right)\, m_i^{z\,[c]} \tag{5b}
\end{equation}
\noindent and 
\begin{equation}
h_i^{x\,\text{eff}\,[c]} = \Gamma. \tag{5c}
\end{equation}

\noindent Here $\vec{m}_i^{[c]} = \langle\vec{\sigma}_i^{[c]}\rangle$, where $< \cdot >$
denotes the thermal average, and the spin
glass order parameter $q (=[q^{[c]}]_{av})$ is given by
\begin{equation}
q = \frac{1}{N} \left[\sum_{i=1}^N \left(m_i^{z\,[c]}\right)^2\right]_{av} .
\end{equation}

The generalized deGennes-Suzuki-Kubo mean field dynamics for the Ising spins \cite{das3} can be represented by a non-linear coupled differential equation (following the refs. \cite{deGennes1963,suzuki, AcharyyaPhysicaA1994,AcharyyaJPhysA1994}):

\begin{equation}
 \frac{d\vec{m}_i^{[c]}}{dt} = - \vec{m}_i^{[c]} +
\tanh\left(\frac{|\vec{h}_i^{\text{eff}\,[c]}|}{T}\right) \frac{\vec{h}_i^{\text{eff}\,[c]}}{\left|\vec{h}_i^{\text{eff}\,[c]}\right|}.
\end{equation}
\noindent This equation can be split into two parts, one for $m^z$ and the other $m^x$.
\begin{equation}
\frac{d m_i^{x\,[c]}}{dt} = - m_i^{x\,[c]} +
\tanh\left( \frac{ \left| \vec{h}_i^{\text{eff}\,[c]} \right| }{T} \right) \cdot \frac{\Gamma} { \left| \vec{h}_i^{\text{eff}\,[c]} \right| } \tag{7a}
\end{equation}

and
\begin{equation}
\frac{d m_i^{z\,[c]}}{dt} = - m_i^{z\,[c]} +
\tanh\left( \frac{ \left| \vec{h}_i^{\text{eff}\,[c]} \right| }{T} \right) \cdot
\frac{ h_i^{z\,\text{eff}\,[c]}  }{ \left| \vec{h}_i^{\text{eff}\,[c]} \right| }
\tag{7b}
\end{equation}

As we consider here a pure quantum annealing scheme, i.e., we set the temperature $T = 0$ for our study, only the transverse field changes with time $t$ according to $ \Gamma(t) = \Gamma(t=0)[1-t/\tau]$, starting from $\Gamma(t=0)=\Gamma_0\ge\Gamma_c$. The value of $\Gamma_c=1$ is considered here for the SK model (see the phase boundary in ref. \cite{das2}). As $T=0$, the `$\tanh$' terms in Eqs. (7a) and (7b) become unity. Here, $t$ is a real physical time (Eq. 7) but it is treated as a discrete iteration counter in the numerical implementation. Hence, for the discrete time ($t$), the above differential equations get further simplified to:
\begin{equation}
m_i^{x\,[c]}(t+1) =
\frac{ \Gamma(t) }{ \left| \vec{h}_i^{\text{eff}\,[c]}(t) \right| }
\tag{7c}
\label{discrete_mx}
\end{equation}
and
\begin{equation}
m_i^{z\,[c]}(t+1) =
\frac{ h_i^{z\,\text{eff}\,[c]} (t) }{ \left| \vec{h}_i^{\text{eff}\,[c]} (t)\right| }.
\tag{7d}
\label{discrete_mz}
\end{equation}

Further, we shall also study the ground state energy of the Edwards-Anderson (EA) model on square lattice employing the same dynamics discussed above. The corresponding Hamiltonian is:
\begin{equation}
    H_{EA} = - \sum_{<ij>} J_{ij}\sigma_i^z\sigma_j^z-\Gamma \sum_i \sigma_i^x.
\end{equation}
Here the first summation is only for nearest neighbors. The distribution of the $J_{ij}$ is taken to be bimodal distribution, i.e., $ \pm 1 $ with equal probability, for most of the study and gaussian distribution (with zero mean and unit variance) for only one instance. The model does not show glass transition at any finite $T$ for $\Gamma=0$. Although for the quantum case, i.e. $T=0$, it does have a critical value of the transverse field ($\Gamma_c$) (see refs. \cite{bkc1981,parisi2024}) where the transition occurs between a spin glass state and a paramagnetic state. The same dynamics (Eqs. (\ref{discrete_mx}) and (\ref{discrete_mz}); with $h_i^{z\,\text{eff}\,[c]}= \sum_{j\in n.n} J_{ij} m_j^{z\,[c]}- \left(1 - q^{[c]}\right)\, m_i^{z\,[c]}$ and $h_i^{x\,\text{eff}\,[c]} = \Gamma$) is employed for the EA model as well in estimating the energy of the ground state. That means we use the same mean-field approximation of the Hamiltonian as done for the SK model and $q$ is computed self-consistently via Eqs. (\ref{discrete_mx}) and (\ref{discrete_mz}) during annealing. The only difference is that the summation of the cooperative part runs only to the four nearest neighbors. However it may be noted that the use of mean-field dynamics can not
be justified pre facto. We will discuss the estimates of ground state
energy value (and justify post facto the use of the annealing
dynamics) in the thermodynamic limit for the EA model in next
section.
%However, it should be noted that by writing a mean-field Hamiltonian for the EA model is a gross oversimplification. We will see the ground state energy value (in the thermodynamic limit) in EA model in next section.  

\section{Numerical Results}
Here, we present the numerical results for the ground state energies of the SK and EA models. The corresponding code is freely available at ref. \cite{code}.
\subsection{SK model}
We numerically solved the above non-linear coupled equation for $m^z$ given by Eq. (\ref{discrete_mz}).  For the entire dynamics $T$ is fixed at zero. That means a pure quantum annealing is considered here. The transverse field is decreased linearly according to $ \Gamma(t) = \Gamma(t=0)[1-t/\tau]$, starting from $\Gamma(t=0)=\Gamma_0=1$ in case of the SK model. Initially, the $m^z$ spins are chosen randomly to be $\pm1$ (discrete) and the $m^x$ be zero, to make the total spin $m=\pm1$. As the dynamics starts, the both $m^z$ and $m^x$ become continuous variables owing to Eqs. (\ref{discrete_mx}) and (\ref{discrete_mz}), and evolve until $t$ becomes $\tau$. At that point, $\Gamma$ is zero and the system becomes purely classical again regaining the integer $m^z$ values. The energy values are then calculated by averaging over the disorder,  $E^0_{N}=- \frac{1}{N}\left[\sum_{i < j} J_{ij}m_i^zm_j^z\right]_{av}$. 

\begin{figure}[tbh]
\hspace*{-0.3cm}
\includegraphics[width=9.5cm]{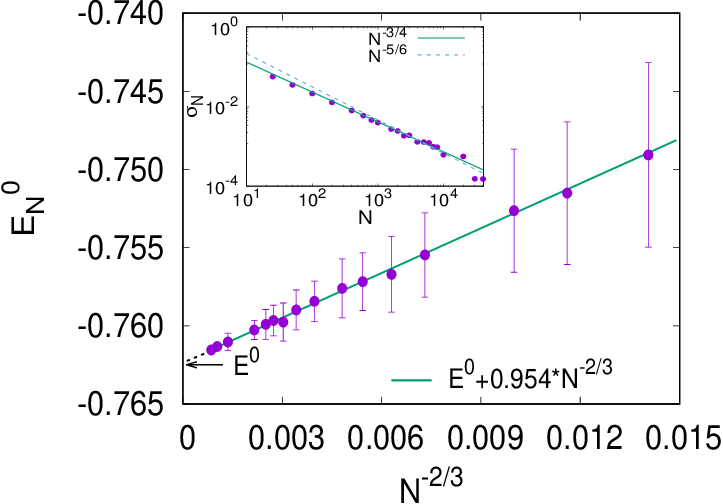}
\caption{Quantum annealing (from quantum paramagnet to zero transverse
field spin glass state) of the SK model: The energy values for given system size are plotted against $N^{-2/3}$.  It shows a scaling $E^{0}_{N}\sim N^{-2/3}$. The ground state energy ($N \rightarrow {\infty}$) is the Parisi value $E^0=-0.7631 \dots$. From the least-square fitting we obtain a ground state energy, $E^0=-0.7623\pm 0.0009$ (considering the exponent to be 2/3; error in extrapolation from fitting with considering maximum error from $N=20000$ onward). The inset shows the variation of the fluctuations $\sigma_N$ of
$E^0_N$. It appears that while for the entire range of $N$ values, $\sigma_N \sim N^{-3/4}$ gives a good fit, for larger values of $N$, $\sigma_N \sim N^{-5/6}$ perhaps gives a better fit. The number of disorder realizations considered are $1000,800,500,300,100,100,80,50,50,50,50,20,24,8,8$ and $4$ for $N=600,800,1000,1600,2000,2500,3000,4000,5000,6000,\\7000,8000,10000,20000,30000$, and $40000$, respectively.}
\label{en0}
\end{figure}
\vskip 0.5cm
For the simulation purpose $\tau$ is chosen to be $5N$. The optimal value of $\tau(=5N)$ corresponds to the onset of energy saturation and this value scales linearly with $N$ (see Fig. 5 of ref. \cite{das1} and Fig. 1 of ref. \cite{das2}). We have simulated the SK model up to a system size of $N=40000$ (number of disorder realization: $8$ for $N=20000,30000$ and $4$ for $N=40000$), and obtained the lowest energy value for $N=40000$ is $E^0_N=-0.7615\pm0.0007$ validating both the usual finite size scaling of the energy $E^0_N-E^0 \sim N^{-2/3}$, with $E^0_{}(N\rightarrow{\infty})=-0.7623\pm0.0009$ (see Fig.\ref{en0}) (compared to the best known estimate $E^0_{}= -0.763166726 \dots$\cite{oppermann}) and finite size scaling of the fluctuation in energy, $\sigma_N\sim N^{-3/4}$ \cite{kobe,das1,das2,das3}.

\subsection{EA model on square lattice}
\subsubsection{\textbf{Annealed ground state energy}}
Similarly, Eq.(\ref{discrete_mz}) for $m^z$ is numerically solved as well in case of an EA model, keeping $T$ fixed at zero and linearly decreasing the transverse field ($\Gamma$ from $\Gamma_0=4$ to $\Gamma=0$) according to above procedure. The same initial spin configurations are also taken here. However, for the EA model, $\tau$ is chosen as $100L^2$ (not $5N$ as in case of the SK model) where the energy values are almost saturated. The energy values are obtained from the equation,  $E^0_{L}=- \frac{1}{L^2}\left[\sum_{<ij>} J_{ij}m_i^zm_j^z\right]_{av}$. 

\begin{figure}[tbh]
\hspace*{-0.6cm}
\includegraphics[width=9cm]{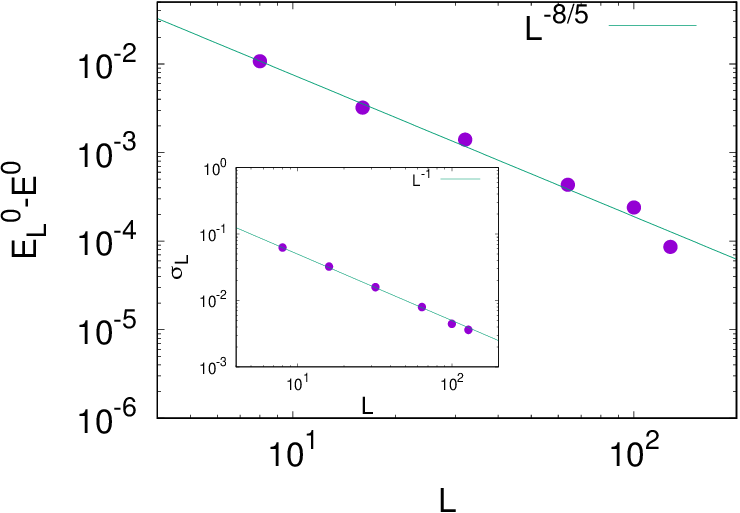}
\caption{Quantum annealing (from quantum paramagnet to zero transverse
field spin glass state) of the EA model on square lattice: In main figure, the finite size scaling of lowest energy values are shown. It shows a scaling $E^{0}_{L}-E^0\sim L^{-8/5}$. The ground state energy ($L \rightarrow {\infty}$) is obtained from the least-square fitting, which is $-1.38\pm0.02$ (considering the exponent to be 8/5). The inset shows the scaling of the fluctuations $\sigma_L$ of
$E^0_L$ which shows a scaling $\sigma_L \sim L^{-1}$. The number of disorder realizations considered are $20000, 10000,2000,1000,$ and $50$ for $L=8,16,32,64,$ and $128$, respectively.}
\label{eneab}
\end{figure}

\begin{figure}[tbh]
\includegraphics[width=9cm]{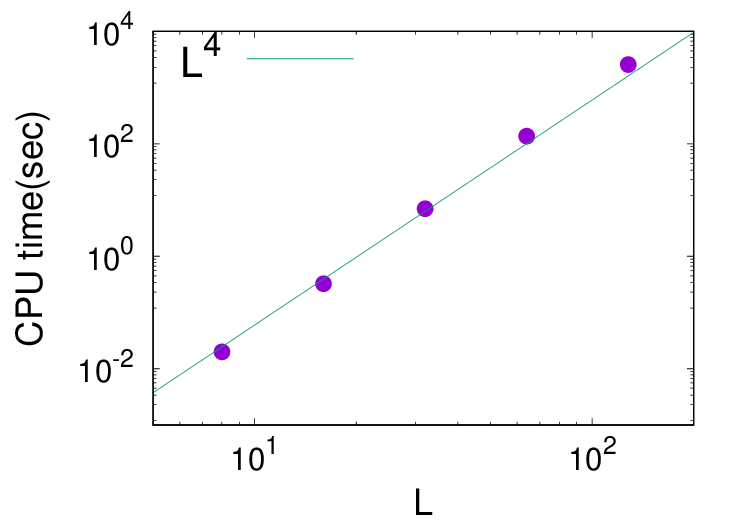}
\caption{The computational cost of the algorithm (for each configurations) in EA model scales as $L^4$ (or $N^2$).}
\label{eab_cpu}
\end{figure}

We have simulated the EA model on a square grid taking the periodic boundary up to a linear system size $L=128$, that is total spins $N=128\times128$. The ground state energy ($L\rightarrow{\infty}$) obtained in this method is $E^0_{}=-1.38\pm0.02$ (compared to the best known results $-1.401938$ \cite{hartmann} and $-1.4024(10)$ \cite{roma}), with a finite size scaling $E^{0}_{L}-E^0\sim L^{-8/5}$ ($N^{-4/5}$). The fluctuation in the energy defined as $\sigma_L= \sqrt{\bigg[{E_L^{0}}^2\bigg]_{av}-\bigg[E^0_L\bigg]_{av}^2}$ scales as $L^{-1}$ ($N^{-1/2}$) (see Fig. \ref{eneab}). Finally, the computation cost scales as $L^4$ (i.e., $N^2$) (see Fig. \ref{eab_cpu}).

%\begin{figure}[tbh]
%\centering
%\includegraphics[width=9.0cm]{qm_eab_open_time_comb.eps}
%\includegraphics[width=8cm]{qm_sd_vsN_new.eps}
%\caption{The total ground state energies of system size $N=8\times8$ (open boundary) are shown for different annealing time $\tau$, (a)$\Gamma=1$ and (b) $\Gamma=2$. For both cases, $E^{tot}_L$ is saturated above $\tau=100\times L^2$.}
%\label{eab_time}
%\end{figure}

\subsubsection{\textbf{Annealed ground state energy for finite $\Gamma$}}
Here both Eqs. (\ref{discrete_mx}) and (\ref{discrete_mz}), for $m^x$ and $m^z$ respectively, are numerically solved, again keeping $T=0$ and linearly decreasing from $\Gamma(t=0)=4$ to $\Gamma (t=\tau)=0.5,1,2$. As we anneal down to a finite value of the transverse in the system, both the average spin components $m^x$ and $m^z$ remain non-vanishing. In other words, both components contribute in the ground state energy, which is the total ground state energy $E^{tot}_{L}=- \frac{1}{L^2}\left[\sum_{<ij>} J_{ij}m_i^zm_j^z-\Gamma\sum_{i}m_i^x\right]_{av}$. 

\begin{figure*}[tbh]
\centering
\includegraphics[width=\textwidth]{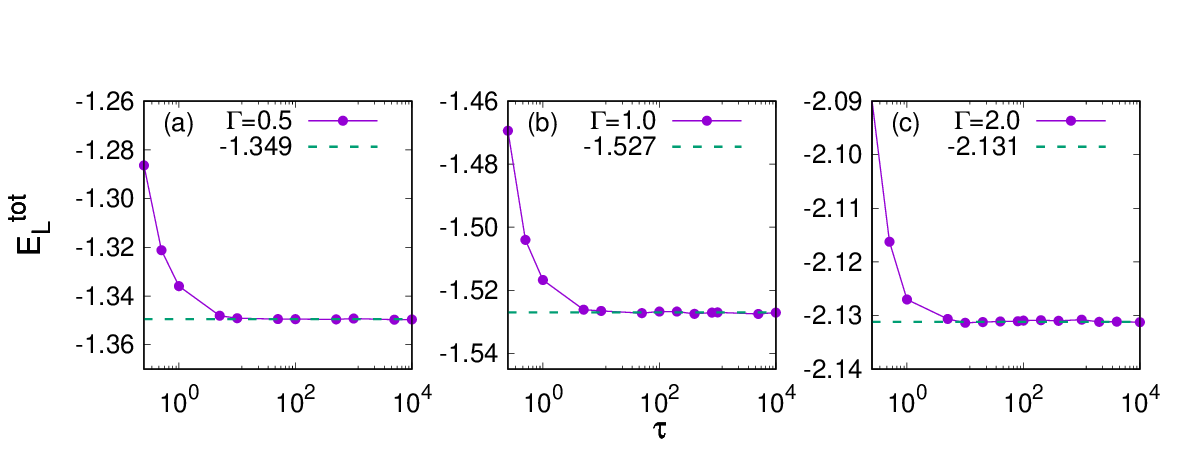}
\caption{Quantum annealing (from quantum paramagnet to finite value of
transverse field) in EA model on square lattice: The saturation of total ground state energy $E^{tot}_{L}$ is shown as a function of annealing time $\tau$ (in units of $L^2$) for different values of $\Gamma$. For all three cases, $E^{tot}_{L}$ saturates quickly as the annealing time increases. For smaller values of the transverse field i.e., in case of (b), $E^{tot}_{L}$ matches well with previously reported result (see ref. \cite{coffrin}),  for $\Gamma=1$, $E^{tot}_{L}=-1.527$. The quantum annealing (at $T = 0$)  is performed for $L = 8$ with open
boundary condition and bimodal distribution of $J_{ij}$. The number of disorder realizations is taken to be $10000$.}
\label{eneab_l8}
\end{figure*}

\begin{figure}[tbh]
\hspace*{-0.5cm}
\includegraphics[width=10cm]{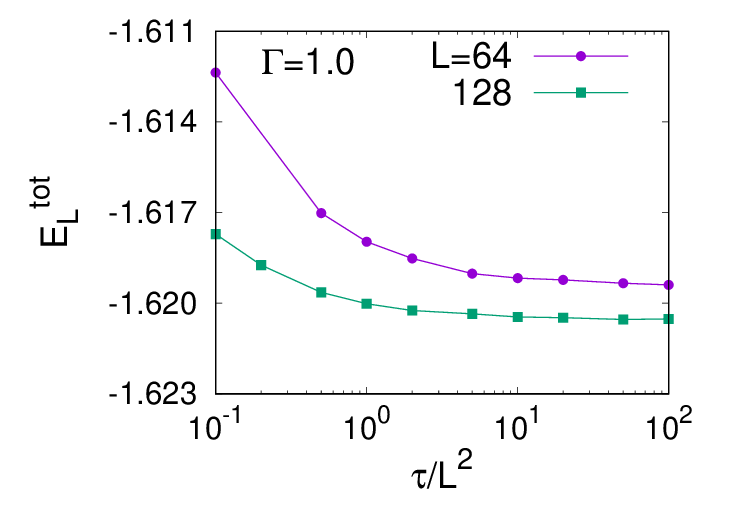}
\caption{2D-EA model on square lattice (with bimodal $J_{ij}$ and OBC): The saturation of total ground state energy $E^{tot}_{L}$ is shown as a function of $\tau/L^2$ for different system sizes ($L=64,128$) for $\Gamma=1$.}
\label{eneab_l128}
\end{figure}

\begin{figure}[tbh]
\includegraphics[width=9cm]{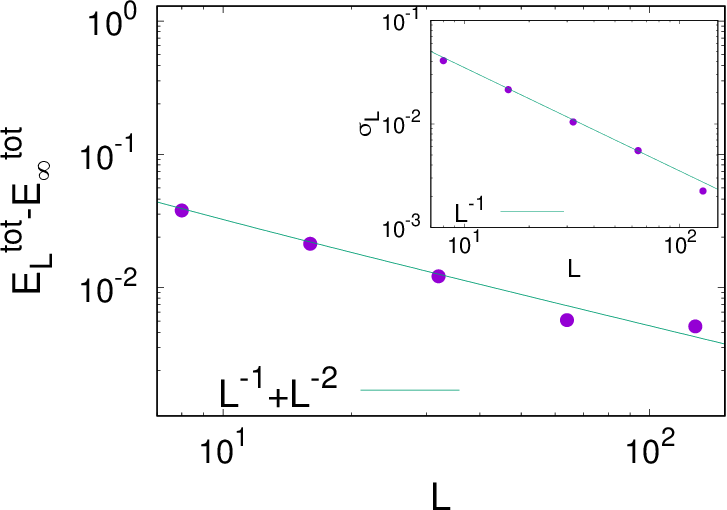}
\caption{Quantum annealing from quantum paramagnet to finite transverse
field ($\Gamma=1$) spin glass state of the EA model on square lattice (with bimodal $J_{ij}$ and OBC): In main figure, the finite size scaling of lowest energy values are shown. It shows a scaling $(E^{tot}_{L}-E^{tot}_{\infty})\sim L^{-1}+L^{-2}$. The ground state energy ($L \rightarrow {\infty}$) is obtained from the least-square fitting, which is $E_{\infty}^{tot}=-1.628\pm0.002$. The inset shows the scaling of the fluctuations $\sigma_L$ of
$E^{tot}_L$ which shows a scaling $\sigma_L \sim L^{-1}$. The number of disorder realizations considered are $100$ for $L=8,16$ and $10$ for $L=32,64,128$.}
\label{qmeab_fss}
\end{figure}

\begin{table}[h]
\centering
\renewcommand{\arraystretch}{2.0}
\setlength{\tabcolsep}{10pt}
\small
\caption{The comparison of the quantum annealing estimate for the
ground state energies of the EA model on square lattice at finite
transverse field ($\Gamma = 1.0$, for bimodal distribution of $J_{ij}$).
The annealing  is performed for $L = 8$, with open boundary conditions
(OBC). Here QA refers to $T = 0$ (fixed) annealing employing  Eqs. (7c) and
(7d), and CA refers to $\Gamma  = 1.0$ (fixed) annealing employing  Eqs. (9a)
and (9b). These estimates can be compared with that reported in ref. \cite{coffrin}, giving  $E^{tot}_8$ ($\Gamma = 1.0$) $\approx -1.52$.}
%\begin{tabular}{|p{3cm}|c|c|}
\begin{tabular}{|l|c|c|}
\hline
\multirow{2}{*}{Method}
& \multicolumn{2}{c|}{$E_8^{\rm tot}$} \\
\cline{2-3}
&
With TAP term
&
Without TAP term \\
\cline{1-3}
QA
&
$-1.52\pm0.04$
&
$-1.53\pm0.04$ \\
\cline{1-3}
CA
&
$-1.52\pm0.04$
&
$-1.53\pm0.04$ \\
\hline
\end{tabular}
\label{tab1}
\end{table}

\begin{table}[h]
\centering
\renewcommand{\arraystretch}{2.0}
\setlength{\tabcolsep}{10pt}
\small      % or \footnotesize, \scriptsize, \tiny
\caption{The comparison of the quantum annealing estimate for the
ground state energies  of the EA model on square lattice at finite
transverse field ($\Gamma = 1.8$, for Gaussian distribution of $J_{ij}$
with zero mean and unit variance). The annealing  is performed for $L =
10$,  with periodic  boundary conditions (PBC). Here QA refers to $T = 0$ (fixed)
annealing employing  Eqs. (7c) and (7d), and CA refers to $\Gamma  =
1.8$ (fixed)  annealing employing  Eqs. (9a) and (9b). These estimates can be
compared with that reported in ref. \cite{pilati2024}, giving  $E^{tot}_{10}$ ($\Gamma =
1.8$) $= -2.10 \pm 0.03$.}

\begin{tabular}{|l|c|c|}
\hline
\multirow{2}{*}{Method}
& \multicolumn{2}{c|}{$E_{10}^{\rm tot}$} \\
\cline{2-3}
&
With TAP term
&
Without TAP term \\
\cline{1-3}
QA
&
$-2.00\pm0.04$
&
$-2.03\pm0.04$ \\
\cline{1-3}
CA
&
$-2.00\pm0.04$
&
$-2.03\pm0.04$ \\
\hline
\end{tabular}
\label{tab2}
\end{table}

We have first considered the EA model (with bimodal $J_{ij}$)  on a
square lattice, taking open boundaries and annealing down (keeping $T = 0$)  to the transverse field values  $\Gamma = 0.5$, $\Gamma = 1$ and $\Gamma = 2$ for $L = 8$ (i.e., $N = 8 \times 8$). The total ground state energy, $E^{tot}_L$ , in this case saturates quickly as the annealing time increases (see Fig. \ref{eneab_l8}). For $\Gamma=0.5$: the saturation occurs at $E^{tot}_L=-1.349$, $\Gamma=1$: $E^{tot}_L=-1.527$, $\Gamma=2$: $E^{tot}_L=-2.131$. In particular, the total ground state energy with open boundaries (for $L=8$) can be compared with a recent paper \cite{coffrin} and both results match reasonably well (see Table I). We also extended the study
of the system, employing ``Classical Annealing" following the discrete
time dynamics
\begin{equation}
   m_i^{x\,[c]}(t+1) =
\tanh\left(
\frac{ \left| \vec{h}_i^{\text{eff}\,[c]}(t) \right| }{T(t)}
\right)
\cdot
\frac{ \Gamma }{ \left| \vec{h}_i^{\text{eff}\,[c]}(t) \right| }  \tag{9a}
\end{equation}
\noindent and

\begin{equation}
   m_i^{z\,[c]}(t+1) =
\tanh\left(
\frac{ \left| \vec{h}_i^{\text{eff}\,[c]} (t)\right| }{T(t)}
\right)
\cdot
\frac{ h_i^{z\,\text{eff}\,[c]} (t) }{ \left| \vec{h}_i^{\text{eff}\,[c]} (t)\right| }, \tag{9b}
\end{equation}
\noindent with $T$ decreasing from $T_0$ to 0, keeping $\Gamma$ value
at the desired low level. 

The results for bimodal $J_{ij}$ with random $\pm 1$ values and with OBC
for  $L = 8$ (as in ref. \cite{coffrin}) and for Gaussian distributed
$J_{ij}$ values and  with PBC for $L = 10$ (as in ref. \cite{pilati2024}) are shown
in Tables \ref{tab1} and \ref{tab2}. The results for CA and QA (also with TAP term and without TAP term; shown in Tables \ref{tab1} and \ref{tab2}) are similar which shows the consistency of both the annealing dynamics in obtaining the total ground state energy. The TAP reaction term is important when the mean field magnitude is very small during the dynamics, while its contribution vanishes at the end of the dynamics. For the SK model, incorporating the TAP reaction term leads to a critical phase diagram that is close to the expected result. However, the same approach does not reproduce a correct critical phase diagram for the 2D-EA model. Nevertheless, it gives similar total ground state energy values with TAP and without TAP term. Furthermore, we have also annealed the quantum EA model on a square lattice of sizes $L=16,32,64,128$ (with bimodal $ J_{ij}$ and open boundary condition), down to $\Gamma=1$ (see Fig. \ref{eneab_l128} for saturation of $E_L^{tot}$ and Fig. \ref{qmeab_fss} for the finite size scaling of $E_L^{tot}$). Fig. \ref{qmeab_fss} gives the estimated value of the ground state energy $E^{tot}_L (\Gamma = 1)$ for the EA model on square lattice (with equal probability bimodal $J_{ij}$) in the limit $L\rightarrow{\infty}$ to be $-1.628\pm 0.002$.

\section{Discussion \& Conclusion}
We have studied quantum annealing of the Sherrington-Kirkpatrick (SK) and
Edwards-Anderson (EA) spin glasses in two dimensions, using the continuous (quantum state)
averaged spins’ deGennes-Suzuki-Kubo quantum Ising mean-field dynamics \cite{das2,das3}. Even though it is already known that the ground state of the mean field SK model can be estimated rather accurately using this approach, it is gratifying to observe the applicability of the deGennes-Suzuki-Kubo mean-field quantum Ising dynamics, albeit
partially, in the context of estimating ground state energy for the
cases of short ranged frustrated systems such as the 2D EA spin glass
as well. 

Particularly the SK spin glass, annealed for up to the size $N=40000$, shows the correct scaling of the residual energy and fluctuations. The extrapolated ground state energy in the thermodynamic limit gives $E^0=-0.7623 \pm 0.0009$ (see Fig. \ref{en0}), which is very close to the theoretical estimate \cite{oppermann} (see also \cite{das2}). A similar estimate of the ground state energy in the case of the EA spin glass on square lattice gives $E^0=-1.38 \pm 0.02$, which is close to the best known estimates of the ground state energies, $E^0=-1.401938$ \cite{hartmann} and $ E^0=-1.4024(10)$ \cite{roma}. Comparatively larger deviation in our annealing estimate for the ground state energy for the (short ranged) EA model is most likely due to the use
of Suzuki-Kubo mean field framework for dynamics. The computational complexities scale as $N^3$ and $N^2$ (or $L^4$) for the SK and EA spin glasses respectively. This scaling follows from the empirically observed linear dependence of the annealing time on $N$, as determined from the saturation of the ground-state energy. A more detailed analysis regarding whether this choice of the annealing time satisfies the adiabaticity criterion relative to the minimum energy gap, is beyond the scope of the present study.

%Still the relatively larger departure for the EA model is due to the short ranged nature of the model's interactions, while the Suzuki-Kubo framework was developed for the mean field dynamics. 

%We have also observed that for Hamiltonian with finite but small values of non-commuting fields, the SUzuki-Kubo dynamics give a good estimate of the ground state energies, as compared to other methods in the same system sizes (see Fig. 5). However, the short ranged nature of the EA spin glass do not match well for its critical point, and therefore not expected to yield close estimates for higher values of the transverse fields. 

We have also observed (as summarized in section III.B.2) that for
annealing down from a quantum para phase (for $\Gamma > \Gamma_c$) to a quantum
spin glass phase (for much smaller $\Gamma$ values), the deGennes-Suzuki-Kubo
dynamics provides fairly good estimates of the ground state
energies of the 2D EA model, in good agreement with recently reported results obtained using different non-annealing numerical methods  \cite{coffrin,pilati2024} (for same system sizes $L = 8$ and $10$; see Tables \ref{tab1} and \ref{tab2}). Later, we
extended our quantum annealing study for the ground state energy of the 2D
EA model annealed down from a quantum para state down to a finite value
of the transverse field ($\Gamma = 1.0$) for system size up to $128
\times 128$ and extrapolated using finite size scaling (see Figs. \ref{eneab_l128} and \ref{qmeab_fss}). The finite size scaling fit (see Fig. \ref{qmeab_fss}) yields an estimate of the ground state energy, $E^{tot}_L (\Gamma =
1)$, for the EA model on square lattice (with equal probability bimodal
$J_{ij}$), in the limit $L\rightarrow{\infty}$, as $-1.628\pm0.002$.

%We have also observed (as summarized in section B2) that for annealing down from a quantum para phase (for $\Gamma > \Gamma_c$) to a quantum spin glass phase (for much smaller $\Gamma$ values), the Suzuki-Kubo dynamics give  good estimates for the ground state energies, as compared to other methods (for the same system sizes; see Fig. \ref{eneab_l128}). However, the critical point estimate for this short ranged 2D EA spin glass does not match well with the best estimates, and therefore  our estimated ground state energies (for relatively higher values of the saturating $\Gamma$ values) are not expected to yield close approximants.

In conclusion, the deGennes-Suzuki-Kubo annealing dynamics for the
continuous averages of the local spin variables, give a fast and
accurate estimate of the ground state energy of the quantum (long
ranged) SK spin glasses. It may be noted that while the mean field
approximation of the dynamics automatically involves the continuous
variables like $m_i^z$ and $m_i^x$ (in say the Eqs. (7)), for quantum
annealing dynamics this is intrinsic and natural.  In general, quantum
annealing performs better due to these natural continuous variations
of $m_i^z$ (even largely ignoring the cooperative interactions and the
resulting energy barriers; without any mean field approximation). This
is because,  in presence of tunable transverse field, having
practically direct control on the $ m_i^x$ magnitude,  the $m_i^z$
gets consequently affected (due to ``tunneling") to satisfy the
$(m_i^z)^2 + (m_i^x)^2 = 1$ constraint. The (cooperatively
interacting) longitudinal spins therefore effectively fail to see the
spin-spin interaction energy barriers directly, as the change in
transverse magnetization will force (by tunneling) the longitudinal
magnetization continuously, leading to the advantages. When extended
to the quantum (and short ranged) EA spin glass, the quantum annealing
framework continues to function reasonably well for the ground state
energy estimation even when annealed to finite (low value of the)
transverse field. This indicates the potential use of this framework
in estimating the ground state energies of the quantum (and
frustrated) many-body systems.

%state  energy of the quantum as well as classical SK spin glasses. When extended to the quantum  EA spin glass, the quantum annealing framework continues to function reasonably well for the ground state energy estimation even when a transverse field is present. This indicate the potential use of this framework in the case of finding ground states of quantum many-body systems.

\section*{Acknowledgments}
We thank Sebastiano Pilati for bringing to our attention the ref. \cite{pilati2024} and the Zenodo link there for their data set. The numerical simulations were performed in HPCC Chandrama, SRM University - AP.

\end{document}